\documentclass{ws-procs9x6}
\newcommand{\beq}{\begin{equation}}
\newcommand{\eeq}{\end{equation}}
\newcommand{\luv}{\Lambda_{UV}}
\newcommand{\lqc}{\Lambda_{QCD}}

\newcommand{\ph}{m_{ph}}
\begin{document}
\title{Fine Tuning in Lattice SU(2) Gluodynamics vs 
Continuum-Theory Constraints}

\author{V.~I. ZAKHAROV}

\address{Max-Planck Institut f\"ur Physik \\
Werner-Heisenberg Institut \\ 
F\"ohringer Ring 6, 80805, Munich\\ 
E-mail: xxz@mppmu.mpg.de}


\maketitle

\abstracts{
Recently, it has been observed that
the non-Abelian action associated
with lattice monopoles and vortices is ultraviolet divergent,
at least at presently available lattices. On the other hand,
the total length of
the monopole trajectories and area of the vortices scale in
physical units. Coexistence of the two different scales,
infrared and ultraviolet, for the same vacuum fluctuations
represents a fine tuning. To check consistency of 
the newly emerging
picture of non--perturbative fluctuations we consider constraints from
the continuum theory on the ultraviolet behaviour of 
the monopoles and vortices. The constraints turn to be satisfied
by the data in a highly non-trivial way. Namely, it is crucial that
the monopoles populate not the whole of the four dimensional
space but a two-dimensional subspace of it.}

\section{Introduction}

By fine tuning one understands usually a particular problem
arising in theory of charged scalar particles. Namely,
expression for the scalar boson mass looks as 
\begin{equation}\label{mass}
m_H^2~=\delta M_{rad}^2-M_0^2~,
\end{equation}
where $\delta M_{rad}^2$ is the radiative correction while $-M_0^2$
is a counter term. The problem is that $\delta M_{rad}^2$
diverges quadratically in the ultraviolet (UV),
$$
\delta M^2_{rad}~\sim\alpha\cdot (const)\int{d^4k\over k^2}~\sim 
\alpha \Lambda_{UV}^2~,
$$
where $\alpha$ is the coupling and $\luv$ is an ultraviolet cut off.
If one uses the Planck mass for the cut off,
$\luv^2~\sim~(10^{19}~GeV)^2$, then to keep the mass of the charged (Higgs)
boson in the $100~GeV$ region one should assume that the counter term is
tuned finely to the value of the radiative correction and this tuning
is readjusted with each order of perturbation theory.

What is outlined above is the standard problem of the Standard Model
but {\it a priori} one would bet that it has nothing to do with 
the vacuum state of the lattice $SU(2)$ theory and with monopoles,
which is a particular kind of the vacuum fluctuations.

However, lattice measurements indicate strongly
(see \cite{anatomy} and references therein) that the monopole mass
is ultraviolet divergent, the same as the radiative correction
in case of a point-like particle:
\beq\label{uvd}
\langle~M(a)_{mon}~\rangle~\sim~{const\over  a}~,
\end{equation}
where $a$ is the lattice spacing  playing the role of an ultraviolet cut off
(in the coordinate space). Moreover, the mass in Eq (\ref{uvd})
is directly related to the excess of the {\it non-Abelian} action
associated with the monopoles.

Let us emphasize that (\ref{uvd}) is a pure phenomenological observation, 
with no theory involved. Which makes it a solid starting point for
a discussion \footnote{To some extent, 
we follow the logic of \cite{gubarev}.}: some objects with property (\ref{uvd})
are certainly there. However, interpretation remains an open question: 
the monopoles themselves are defined  not in terms of  
original Yang-Mills fields but rather in terms of projected fields,
for reviews see, e.g., \cite{reviews}.
Namely, for a given configuration on replaces $SU(2)$ by the ``closest''
$U(1)$ field configuration.( Through projecting to the
closest $Z_2$ configuration one can define vortices, for review see 
\cite{greensite}). 
This is a well defined algorithm and
the results like (\ref{uvd}) are unique.
One could define, however, monopoles in 
a different way and then their properties would change, generally speaking. 

Under the circumstances, we feel that the following strategy could
be appropriate. We will assume that, through the projection,
one detects in fact --to some accuracy and on average-- 
gauge invariant objects \footnote{
Such a possibility is noticed, in particular, in Ref \cite{hart}.}. 
Then one can look for other gauge invariant characteristics
and, indeed, there is accumulating evidence that there exist
further $SU(2)$ invariant properties of the monopoles, see, in
particular, \cite{muller,boyko1,kovalenko,boyko2}.
Still, this evidence is pure numerical and, at some time, 
it would be desirable to switch into
the language of the continuum theory. 
Our point here is that the very fact of appearance of
the {\it ultraviolet} divergence (\ref{uvd}) allows 
to establish strong constraints from the continuum theory.
In particular, the
asymptotic freedom does not allow new particles,
in apparent contradiction with (\ref{uvd}). Closer examination 
reveals, however, that according to the data the monopoles
are not ordinary particles indeed. Since they occupy not
the whole of the $d=4$ (Euclidean) space but rather
a $d=2$ subspace of it, for a recent review see an accompanying
paper by the same author \cite{vz1}. We will argue that
existence of such `branes' is consistent with the asymptotic freedom.
A posteriori, one can say that this is a unique way to reconcile
(\ref{uvd}) with the asymptotic freedom. The theory of the branes themselves
is lacking, however.
 
\section{$SU(2)$ fine tuning, seen on the lattice}

As is mentioned in the Introduction, monopoles in $SU(2)$ are defined 
through a projection on $U(1)$ fields, for review see 
\cite{reviews,greensite}. 
In more details, one starts with 
generating a representative set of vacuum configurations of the original
Yang-Mills theory with the standard action. Then each configuration --
and this is the central point -- is projected into the
closest configuration of $U(1)$ fields. The projection itself is in two steps.
First,  one uses gauge invariance to minimize, over the whole lattice 
the functional
\beq
R~=~\sum_{links}\big[(A_{\mu}^1)^2~+~(A_{\mu}^2)^2\big]~~,
\end{equation}
where $A_{\mu}^i$ is the gauge potential and $i (i=1,2,3)$   
are color indices. The meaning of minimizing $R$ is that `charged' fields are
minimized. This fixation of the gauge does not change
physics, of course. However, at the next step, which is the projection itself, 
one sends $A_{\mu}^{1,2}$ to
zero generating in this way effective $U(1)$ fields, $\bar{A}^3_{\mu}$
\footnote{The hypothesis behind is, of course, Abelian dominance in the
infrared. Consistency of this hypothesis can be checked. For a recent and 
amusing example of the Abelian dominance in the infrared see \cite{morozov}.}.

Finally,  the monopole current, $j_{\nu}^{mon}$ 
is related to violations of the
Bianchi identities in terms of the projected fields:
\beq\label{current}
\partial_{\mu}\tilde{F}_{\mu\nu}~=~j_{\nu}^{mon}~~,
\end{equation}
where $F_{\mu\nu}$ now is the field strength tensor constructed
on the projected fields $\bar{A}^3_{\mu}$ and $\tilde{F}_{\mu\nu}$ 
is dual to $F_{\mu\nu}$.
A non-vanishing current (\ref{current}) implies 
projected fields to be singular in the continuum limit.
However, all the singularities are regularized by the lattice and
the expression (\ref{current}) is well defined \cite{degrand}.

The vortices are defined
in terms of projected, or closest $Z_2$ configurations which
are matrices $\pm I$ ascribed to each link. The vortices are  unification
of all the negative plaquettes evaluated on the $Z_2$ projection. The
corresponding surfaces are closed by definition, as boundary of a boundary. 

An analysis of this type ends up with a net of monopole trajectories 
or central vortices
for each original non-Abelian field configuration.
Theoretical task is then to interpret the data on the monopole 
and vortex clusters. Because of the use of the projection
this task turns to be very difficult.

 Phenomenologically,
both lattice monopoles and vortices exhibit remarkable properties.
Which we will briefly summarize here. Note that  we present a simplified
picture, emphasizing only the main features of the data (as we 
appreciate them). For details, one is to consult  the original papers. 

\begin{itemize}
\item
There is an excess of non-Abelian action associated with 
the monopoles \cite{anatomy} which  can be memorized as:
\beq\label{maction}
<S_{mon}>~\approx~\ln 7\cdot{L\over a}~~,
\end{equation}
where $L$ is the length of the monopole trajectory. The overall constant in
(\ref{maction}) is actually poorly known. We put it equal to 
``${\ln 7 }$'' 
since this value 
does not contradict the data, on one hand, and would have a 
simple theoretical interpretation, on the other. 

\item For each field configuration, there exists a single 
percolating cluster
which extends through the whole of the lattice. The length of the
corresponding trajectory per unit volume does not depend on $a$ and scales
in the physical units, see \cite{muller} and references therein. 
The observation can be formulated also in
terms of the probability of a given link on the (dual) lattice to belong
to the percolating cluster: 
\beq\label{link}
\theta (link)~\sim~(a\cdot\lqc)^3~~,
\end{equation}
for all the values of $a$ tested. 

\item In case of vortices, the total area scales in the physical units,
see \cite{greensite} and references therein. Numerically \cite{kovalenko}:
\beq \label{rho} 
A_{vort}~\approx~24(fm)^{-2}\cdot V_4~,
\end{equation}
where $A_{vort}$ is the total area of the vortices in the lattice
volume $V_4$.
One can rewrite (\ref{rho}) in terms of probability for a particular
plaquette to belong to the percolating vortex:
\beq\label{plaqe}
\theta (plaq)~\sim~(a\cdot\lqc)^2~~.
\end{equation}

\item The non-Abelian action associated with the vortices is
ultraviolet divergent \cite{kovalenko}:
\beq\label{vaction}
<S_{vort}>~~\approx~0.54\cdot {A\over a^2}~~.
\end{equation}

\end{itemize}
 It is worth emphasizing that all the properties (\ref{maction}), 
(\ref{link}),
(\ref{plaqe}) and (\ref{vaction}) are perfectly gauge invariant.
Thus, the data suggest that, through projections, one detects gauge invariant 
objects.

\section{Fine tuning: well understood examples}

Since singular non-Abelian fields have an infinite action 
in the continuum limit,
common wisdom tells us that such fields would drop off
by themselves and, therefore, one would assume that the singularity 
(\ref{current}) arises as an artifact of the projection.
Observations on the monopoles and vortices summarized above imply 
that something is missing in this 
standard logic \cite{vz}. 
Generically, coexistence of the ultraviolet and
infrared scales can be called ``fine tuning''. (Relation to the fine tuning of
the standard model will be clarified later.)
To orient ourselves in the problem, we will start with reviewing cases
when the fine tuning is well understood.

\subsection{Free particle}\label{subsec:fp}

Consider first a free particle with the  classical action \footnote{
This subsection is mostly a text-book material, see, e.g., \cite{ambjorn}.}:
\beq\label{cl}
S_{cl}~=~M(a)\cdot L~~,
\end{equation}
where $M(a)$ is a mass parameter and we reserved for its possible dependence
on the lattice spacing, while $L$ is 
a length of a trajectory of the particle,
everything in the Euclidean space.

Furthermore, define a propagator as a path integral:
\beq\label{propagator}
D(x_i,x_f;a)~=~\sum_{paths}\exp(~-S_{cl})~~,
\end{equation}
where $x_{i,f}$ are the end points of trajectories.
The summation in (\ref{propagator}) can be performed explicitly.
In the momentum space:
\beq
D(p;a)~=~{1\over c^2a^2}D_{free}(m_{ph}^2)~~,
\end{equation}
where $c$ is a constant depending on details of the ultraviolet regularization
and $\ph$ is the propagating mass. The relation of $\ph$ to the 
bare mass $M(a)$ introduced in (\ref{cl}) is as follows:
\beq\label{ph}
\ph^2~=~{8\over a}\big(M(a)~-~{\ln 8\over a}\big)~~,
\end{equation}
where the constants in front of the ultraviolet factors 
(i.e., inverse powers of $a$) are in fact 
regularization dependent and  hereafter
we have in mind hyper-cubic lattice.
Eq (\ref{ph}) demonstrates that to keep the physical mass fixed, 
i.e. independent
of $a$, one should tune the bare mass to a pure geometrical factor. 

One could proceed further and 
consider interaction as well.
Here, we will use this approach only to derive
a useful relation for the vacuum expectation value of the corresponding
scalar field squared \cite{vz,maxim}.
Namely, the average value of the length of the particles trajectories
in the vacuum is given by:
\beq
\langle~L~\rangle~=~{\partial\over \partial M}\ln~Z~~,
\end{equation}
where $Z$ is the partition function.
Moreover, one can replace:
\beq
{\partial\over \partial M}~\rightarrow~{8\over a}{\partial\over 
\partial \ph^2}~.
\end{equation}
The derivative with respect to $\ph^2$, on the other hand, is related
to the vacuum expectation of the $|\phi|^2$
where $\phi$ is a (complex) scalar field entering the standard formulation
of field theory. Indeed, the standard Lagrangian contains a term
$\ph^2|\phi|^2$. 

Finally,
\beq\label{relation}
V_4\langle~0|~|\phi|^2~|0~\rangle~=~{a\over 8}\langle ~L~\rangle~~,
\end{equation}
where $\langle ~L~\rangle$ is average length of trajectory in
the volume $V_4$.
Eq (\ref{relation}) relates quantities entering the standard and 
polymer representations of theory of a scalar field. For us, it is important
that the lengths of the monopole trajectories are directly measurable.

\subsection{Lattice $U(1)$}

Lattice $U(1)$ , see, e.g., \cite{polyakov}, is actually close 
to the case of free particle
just considered. A new point is that 
$M(a)$ is now calculable as energy of the magnetic field:
\beq\label{mon}
M(a)_{mon}~=~{1\over 8\pi}\int {\bf H}^2d^3r~\sim~{const\over e^2 a}~~,
\end{equation} 
where one has to introduce an ultraviolet cut off, $a$ since $H\sim 1/r^2$ 
and the integral diverges at small distances.  Note also that we kept 
explicit dependence
on the electric charge $e$ which is due to the Dirac quantization
condition. Finally, and might be most noteworthy,
Eq (\ref{mon}) does not contain contribution of the Dirac string.
This is a privilege of the lattice regularization (for more details see,
e.g., \cite{alive}). 

Note that upon substituting (\ref{mon}) into (\ref{ph}) we reproduce in fact
(\ref{mass}).
Now, if one tunes $e^2$ in such a way that $\ph=0$, where $\ph$
is defined in (\ref{ph}) the monopoles condense. This is confirmed 
by the lattice data \cite{shiba}.

\subsection{Percolation}

Percolation is a common notion in papers on the monopoles
(for a review of percolation theory see, e.g.,
\cite{fortunato}). In most cases it is related to existence of 
an infinite cluster of monopoles,
see \cite{hart} and references therein.
Phenomenologically the percolating cluster
is very important since the
confining potential for external heavy quarks is 
entirely due to this infinite cluster
while finite clusters do not confine. 

Uncorrelated percolation is the simplest kind
of percolation. 
In this case, one introduces
a probability $p, p<1,$ for a link to be ``open''
and this probability does not depend on the neighbors. In our case, an open
link would correspond to a link belonging to a monopole trajectory
\footnote{Monopoles are defined as end points of the Dirac strings 
and occupy centers of lattice cubes. Alternatively, one can say that
on the dual lattice monopoles
occupy sites and the monopole trajectories are built up on 
the links on the dual
lattice. In most cases, we do not mention that it is the dual lattice which
is implied in fact.}. The probability to find a connected
trajectory of length $L$ is given by
\beq\label{w}
W(L)~=~p^{L/a}\cdot N_L~~,
\end{equation}
where $L/a$ is the number of steps and $N_L$ is the number of
various trajectories of the same length $L$. 
Moreover,
\beq\label{nl}
N_L~=~8^{L/a}~.
\end{equation}
Indeed, the monopoles occupy centers of cubes and at each
step the trajectory can be continued to a neighboring cube.
There are 8 such cubes for $d=4$ \footnote{For charged particles, 
which we are considering, 
the factor 8 in Eq (\ref{nl}) is to be replaced by the factor of 7.
Indeed, if one and the same link is covered by a trajectory 
in the both directions, then the link does not belong to a trajectory at all.
In the field theoretical language
this cancellation corresponds to the fact particle and 
anti-particle have opposite charges. For simplicity of presentation we will
keep Eq (\ref{nl}) without change}.
Note that uncorrelated percolation is equivalent to a free field theory.
Indeed by identification
$$
M(a)~=~{\ln p/~ a}~~.
$$
the factor $p^{L/a}$ in Eq (\ref{nl}) reduces to the action factor for a free
particle while $N_L$ represents the entropy.

\subsection{Supercritical phase of percolation}
 
Clearly, there is a critical value of $p,~ p=p_c,$ when 
 any length $L$
is not suppressed. This is the point of phase transition to percolation.
In the supercritical phase, $p>p_c$, there always exists a single
infinite percolating cluster \cite{fortunato}. 
Most interesting, if $(p-p_c)\ll 1 $ 
the probability for a link to belong to the percolating cluster is also small:
\beq\label{theta}
\theta (p)~\sim~(p-p_c)^{\alpha}~~,
\end{equation}
where the critical exponent $0<\alpha<1$. In other words, the supercritical
phase can be consistently treated as far as $p-p_c\ll 1$ \cite{fortunato}.

A simple effective action to describe the percolating monopole trajectory
was proposed recently in \cite{ishiguro}:
\beq\label{effective}
S_{eff}~=~-\mu\cdot L_{perc}~+~\gamma{L^2_{perc}\over V_4}~~,
\end{equation}
where $\mu,\gamma$ are, generally speaking, functions of $a,\lqc$.
Note that $\mu$ is positive so that in the (formal) limit of $V_4\to \infty$
the action (\ref{effective}) corresponds to a single {\it tachyonic} mode.
Because of the tachyonic mode the theory is actually not defined at all
without the $L^2$ term in (\ref{effective}).

For a finite volume one can readily determine
the average length of the percolating cluster:
\beq\label{central}
<L_{perc}>~=~{\mu\over 2\gamma} V_4~~,
\end{equation}
and  adjust the coefficients $\mu,\gamma$ to reproduce the
observed density of percolating monopoles. Moreover, action (\ref{effective})
can describe, for a finite volume, also fluctuations of $L_{perc}$ around 
its central value (\ref{central}) \cite{ishiguro}.

The $L^2$ term in (\ref{effective}) corresponds in fact to the
specific heat in the language of thermodynamics and could be postulated 
on general grounds \footnote{The remark is due to L. Stodolsky,
for more details see \cite{wosiek}.}. Basing on the general arguments one
can expect that the coefficient $\gamma$ tends to zero as $p$ approaches
$p_{c}$ from above.

\subsection{Lattice $Z_2$ gauge theory}

Because of space considerations, we cannot go into details of the  lattice
$Z_2$ gauge theory. Roughly speaking,  the mechanism of the phase transition
to percolation is similar to the $U(1)$ case, with replacement
of trajectories with closed $d=2$ surfaces.namely, both the action
and entropy factor are divergent in ultraviolet as exponents
of $A/a^2$ and can be tuned to each other by choice of the coupling.

\section{Data vs. theoretical constraints}
\subsection{Asymptotic freedom and counting degrees of freedom}

We see that both in case of $U(1)$ and $Z_2$ gauge theories confinement
can be ensured by tuning the corresponding couplings. 
Moreover, the idea that the $SU(2)$ confinement is similar
to the $U(1),Z_2$ cases was the driving force to finally discover
the amusing properties of the monopoles and vortices in the
$SU(2)$ case. Also, Eqs. (\ref{link}), (\ref{plaqe}) 
look typical for a percolating system
in the supercritical phase, see Eq. (\ref{theta}).

Nevertheless, there are accumulating arguments that the
fine tuning in the $SU(2)$ case is to be different. 
On the theoretical side and to begin with, 
there is no coupling to tune since it 
is running. Moreover, the 
monopoles and vortices are not intrinsic to the full $SU(2)$,
but rather to its subgroups,
that is $U(1)$ and $Z_2$. Namely, there is no topological definition of
monopoles (vortices) in the full $SU(2)$ which would imply lower
bounds on the action of the topologically non-trivial fluctuations. 
As a result, there is no answer to the question why the monopole or
vortex action cannot go down \cite{alive}. If it were allowed, however,
monopole and vortices would have packed the whole of the lattice.  

Now, that it has been revealed that the monopoles and vortices
are associated with divergent actions, see Eqs. 
(\ref{maction}) and (\ref{vaction}) the problems become even more acute.
Indeed, Eq. (\ref{maction}) looks exactly the same as for a point-like 
particle but it is clear that in an asymptotically free theory we are not 
allowed to add new particles.

Let us look closer, what the actual constraints are.
Consider the `cosmological constant' , that is density of vacuum energy:
\beq\label{cosm}
\epsilon_{vac}~\approx~\sum_{{\bf k}}{\omega ({\bf k})/~2}~~.
\end{equation}
In an asymptotically free theory
Eq (\ref{cosm}) should be a valid approximation. The sum (\ref{cosm})
diverges in the ultraviolet as $a^{-4}$. The coefficient in front of
the divergence depends on the number of
degrees of freedom. Moreover, all the divergences are regularized by 
the lattice, so that Eq (\ref{cosm}) is well defined and can be used
to predict the average value of the plaquette action $\langle P\rangle$
on the lattice:
$\langle P\rangle$:
\beq\label{cg}
\langle 1- P\rangle_{pert}~\approx~{c_G\over a^4}~~,
\end{equation}
where the coefficient $c_G$ is known,
for further
details and references see \cite{gc}.

However, this constraint is formulated in terms of the original 
gluonic fields.
To appreciate the meaning of the constraint in terms of the monopole 
trajectories,
which are our basic observables now, notice that an uncorrelated
percolation is equivalent to a `free particle' in the tachyonic mode,
see Sect. 3.4.
The crucial point is then that in the percolation picture
there is nothing happening to the {\it total} monopole density at $p=p_c$.
Indeed the total density of `open' links is simply:
\beq\label{total}
\rho_{tot}^{perc}~=~p\cdot {1\over a^3}~,
\end{equation}
and there is no discontinuity or non-analyticity at $p=p_c$. 
It is only the density of the infinite cluster which exhibits a threshold
behaviour (\ref{theta}). 

Existence of 
the percolating cluster is crucial for the confinement and that is why
one usually concentrates on $\rho_{perc}$. However, now we come to the
conclusion that from the theoretical point of view it is the
total  monopole density  which is constrained by the asymptotic freedom.
If the density of the percolating cluster is vanishing at the point
of the phase transition (see (\ref{theta})) where the total
density (\ref{total}) goes to? Clearly, to finite clusters. 
Thus, we should address theory of finite clusters.

 \subsection{Monopoles as a novel probe of short distances}

Let us consider first free particle case.
Then the simplest Feynman graph is a vacuum
loop. Usually, text-books say that this graph is not observable since
all the amplitudes are normalized to the vacuum-to-vacuum transition.
Similarly, one is usually saying that the vacuum energy is normalized to
zero by definition, However, once the lattice regularization is introduced,
the `cosmological constant' turns directly observable.
The same is true for the vacuum-to-vacuum
transition. For free particles, vacuum loops are directly observable and
one can try to evaluate them theoretically \cite{maxim}.

Moreover, the calculations are in fact straightforward. 
Since it is the trajectories that are directly observable one is invited to use
the polymer representation of field theory, see Sect. 2.1.
There are two basic properties of finite clusters which can be
predicted starting from the assumption that
the monopoles
at short distances can be treated as free particles \cite{maxim}. 
First, the spectrum of the finite clusters as
function of their length is given by \footnote{The result can be actually 
read off from the equations derived, e.g., in Ref. \cite{polyakov1}.}: 
\beq\label{spectrum}
N(L)~=~{const\over L^{d/2+1}}~=~{const\over L^3}~~,
\end{equation}
where we substituted the number of dimensions of the space $d=4$
\footnote{The Coulomb-like interaction in $d=4$ leaves
actually (\ref{spectrum}) and (\ref{radius}) unchanged \cite{maxim}.}.
Second, radius of a cluster
is predicted to be:
\beq\label{radius}
R(L)~\sim~\sqrt{L\cdot a}~~.
\end{equation}

The predictions (\ref{spectrum}) and (\ref{radius}) are in perfect 
agreement with the dat\cite{hart,boyko2}. 
Eq (\ref{spectrum}) is especially remarkable for the fact that
the spectrum depends explicitly on the number
of dimensions of space-time. Thus, at short distances the
monopoles move as free particles in $d=4$ space.

\subsection{Conspiracy of the monopoles and vortices}

The fact that the predictions (\ref{spectrum}) and (\ref{radius})
agree with the data, at first sight, is in contradiction with what
we said earlier on counting degrees of freedom at short
distances. However, Eqs (\ref{spectrum}), (\ref{radius}) do not yet
allow for a direct comparison with expectations based
on asymptotic freedom. Consider therefore the vacuum expectation value
(\ref{relation}). Clearly, the asymptotic freedom requires that
\beq\label{modest}
\langle~0|~|\phi|^2~|0~\rangle~\sim~\lqc^2~~,
\end{equation}
while any ultraviolet divergence in this vacuum expectation value would
imply that monopoles are to be introduced on equal footing with
the gluons and this  is not allowed.

Constraint (\ref{modest}) can be turned into a constraint on 
monopole densities defined as:
\beq
L_{mon}~=~L_{perc}~+~L_{fin}~\equiv~\rho_{perc}V_4~+~\rho_{fin}V_4~~,
\end{equation} 
where $L_{perc}$ and $L_{fin}$ are the lengths of the trajectories 
belonging to the percolating and finite clusters, respectively.
Thus, Eq. (\ref{modest}) implies:
\beq\label{constraint}
\rho_{fin}~\le~ {const^{'}\over a}~~.
\end{equation}
It is most amusing that the constraint (\ref{constraint}) is
satisfied by the data! The first indication to the $1/a$ behaviour 
of the total density of the monopoles was obtained in Ref \cite{muller}
and is now confirmed in Ref \cite{boyko2}. 

It is easy now to figure out the geometrical meaning of the relation
(\ref{constraint}) \cite{boyko2}. Clearly, monopoles
are to be associated with a $d=2$ subspace of the whole $d=4$ space.
Taken as a prediction, this sounds bizarre. Unfortunately, no prediction
of this kind was done (and our analysis is post hoc). 
The fact that the vortices are populated
by monopoles is rather known empirically \cite{giedt,kovalenko}.
Thus, there exists a long-range correlation between finite clusters.
As a result, the constraint (\ref{modest}) gets
satisfied by the data. It is amusing that that at short distances  monopoles
are `primary' object and the vortices span on the
trajectories, see discussion around Eq. (\ref{spectrum}). At large distances,
to the contrary, the vortices are `primary' and monopoles are `secondary'.
Striking differences between the ultraviolet and infrared behaviour
are well known in case of random walks, for a text-book presentation see,
e.g., \cite{ambjorn}, while data on the monopole clusters can be found in
Ref. \cite{boyko1}. For the surfaces, we believe, the topic is quite new. 

To reiterate the central point: 
in case of uncorrelated  percolation we would get
\beq\label{qd}
\langle~0|~|\phi|^2~|0~\rangle_{perc}~\sim~p\cdot a^{-2}~~
\end{equation}
recovering the standard quadratic divergence which plagues
theory of scalar elementary particles. As we emphasized above,
(\ref{qd}) is not allowed by the asymptotic freedom. For similar reasons
we could not allow for percolation of fine tuned surfaces which is
relevant to the pure $Z_2$ case. On the other hand, percolation of `branes',
i.e. of $d=2$ surfaces populated with monopoles is allowed by the constraint
(\ref{modest}) and is observed on the lattice.  

\subsection{Gauge-invariant approaches}

As is mentioned a few times above, no gauge invariant description
of the vortices has been developed. Let us still mention that
it seems natural to consider determinant $D$ constructed on three
independent color magnetic fields, for a given choice of the `time' slice:
\beq\label{determinant}
D({\bf x})~=~\Vert H^a_i({\bf x})\Vert ~~,
\end{equation}
where $a=1,2,3$ is a color index and $i=1,2,3$ is a space index.
Zeros of the determinant (\ref{determinant}) define a two-dimensional
surface. One could try to identify these surfaces with the vortices
detected on the lattice. There is a phenomenological support for 
such an identification
since the vortices are indeed `thin' and this corresponds 
to color magnetic field spread tangentially on surfaces. Monopoles
would correspond to a zero of second order and even more asymmetric field
(at short distances).

Another way of searching for two-dimensional surfaces in an explicitly
gauge-invariant fashion is to study properties of the Dirac operator,
for details 
see \cite{thacker} and references therein. Moreover, vanishing of
the determinant (\ref{determinant}) means changing of a left-handed
configuration of the magnetic fields into the right-handed and vice versa.
Thus, surfaces associated with degeneracy of the magnetic fields could well
be detected through studies of the chiral modes and the vortices discussed
here could be identical or closely related to the surfaces found in Ref. 
\cite{thacker}.

Finally, let us mention that combining Eq. (\ref{relation}) with
the data on the monopole density in finite clusters one can find:
\beq\label{result}
\langle \vert \phi\vert^2\rangle~\approx~0.8 ~(fm)^{-2}~~.
\end{equation}
The result (\ref{result}) is perfectly gauge invariant. Moreover, 
the condensate has dimension 2 and one can wonder, what kind of condensate
in terms of the fundamental gluon fields can be behind (\ref{result}).
This question could have been considered as a serious objection
to (\ref{result}) since, naively, the simplest gluon
condensate, $\langle (G_{\mu\nu}^a)^2 \rangle$ has dimension 4.
However, it was demonstrated recently that dimension-two condensate
can well be defined in terms of gluon fields \cite{stodolsky}.

\section{Conclusions}

We have argued that the fine tuning exhibited by the lattice data
on $SU(2)$ gluodynamics appears to be a novel phenomenon.
Namely, the density of the percolating monopole and lattice clusters 
are similar  to well known examples of the lattice $U(1)$ and $Z_2$
gauge theories near the phase transition. 
However, in the latter cases there exist elementary monopoles (vortices).
As a result, vacuum expectation values of the corresponding fields
are ultraviolet divergent. Such a divergences would be inconsistent
with the asymptotic freedom in the non-Abelian case.
And this constraint, at first sight, rules out an
ultraviolet divergence in the monopole mass (see Eq (\ref{uvd}))  
which is similar to the point-like particle.
The way out is suggested by the data. Namely, the monopoles
populate not the whole $d=4$ space but rather a $d=2$ subspace of it.
Existence of these new objects (``branes'') is consistent with 
the continuum theory. Which adds credibility to the lattice evidence.

Note that, combining Eqs. (\ref{rho}) and (\ref{vaction})
we find that the contribution of the vortices into the average plaquette action
is of order
\beq\label{renormalon}
\langle 1- P\rangle_{vort}~\approx ~0.1 ~GeV^2~a^{-2}~~.
\end{equation}
This contribution matches the so called ultraviolet renormalon,
for discussions see, in particular, \cite{gc,grunberg}. We 
hope to come back to this issue in
a separate note.

\section*{Acknowledgements}

This mini-review is based on a talk presented to the Pomeranchuk 
memorial Conference held at the Moscow Institutes for Physics and Engineering
(MIPhE) in January 2003. The Conference was a moving event for me.
I met first Isaak Yakovlevich Pomeranchuk while a student of MIPhE 
(and he was teaching at the Institute). His devotion to physics and personality
crucially influenced physicists around him, not even necessarily very
close to him.  The pain caused by his
untimely death in 1966 is still alive for everyone of this circle.
I am very thankful to the organizers of the Conference for their efforts.

Detailed discussions and collaborations with V.G. Bornyakov, P.Yu. Boyko,
M.N. Chernodub, F.V. Gubarev, V.A. Kovalenko, M.I. Polikarpov 
and  S.N. Syritsyn are gratefully acknowledged. I am also thankful to 
J. Greensite, P. Nieuwenheuzen, T. Suzuki 
and L. Stodolsky for interesting discussions.

\end{document}